\documentclass[aps,prl,reprint,twocolumn,groupedaddress]{revtex4-2}

\usepackage{hyperref}
\usepackage{amsmath}
\usepackage{amsfonts}
\usepackage{amssymb}
\usepackage{graphicx}
\usepackage{empheq}
\usepackage{diagbox}
\usepackage{tikz,xcolor}

\usepackage{xr}
\usepackage{float}

\definecolor{linkcolour}{HTML}{000066}	
\hypersetup{colorlinks,breaklinks,
	urlcolor=linkcolour, 
	linkcolor=linkcolour,
	citecolor=linkcolour}

\definecolor{lime}{HTML}{A6CE39}
\DeclareRobustCommand{\orcidicon}{
	\begin{tikzpicture}
		\draw[lime, fill=lime] (0,0) 
		circle [radius=0.16] 
		node[white] {{\fontfamily{qag}\selectfont \tiny ID}};
		\draw[white, fill=white] (-0.0625,0.095) 
		circle [radius=0.007];
	\end{tikzpicture}
	\hspace{-2mm}
}

\makeatletter
\def\maketitle{
	\@author@finish
	\title@column\titleblock@produce
	\suppressfloats[t]}
\makeatother

\newcommand{\orcidauthorNB}{\href{https://orcid.org/0000-0002-1554-3820}{\orcidicon}} 
\newcommand{\orcidauthorMRF}{\href{https://orcid.org/0000-0001-5864-9636}{\orcidicon}} 

\hypersetup{colorlinks=true, urlcolor=blue, linkcolor=blue, citecolor=blue,plainpages=false} 

\newcommand{\tr}{\mbox{tr}}

\newcommand{\partialdiff}[2]{\partial_{#2}#1}

\newcommand{\res}[2]{\mathop{\mathrm{Res}}_{\omega = #1}#2}
\newcommand{\real}[1]{\mathrm{Re}#1}
\newcommand{\imag}[1]{\mathrm{Im}#1}
\newcommand{\dee}{\mathrm{d}}

\newcommand{\her}[1]{\mathrm{Her}#1}
\newcommand{\sher}[1]{\mathrm{SHer}#1}
\usepackage{datetime}
\newdate{date}{21}{08}{2024}

\begin{document}

\preprint{APS/123-QED}

\title{Perturbing scattering resonances in non-Hermitian systems: a generalized Wigner-Smith operator formulation} 

\author{Niall Byrnes\orcidauthorNB$^1$, and Matthew R. Foreman\orcidauthorMRF$^{1,2}$}
\email[]{matthew.foreman@ntu.edu.sg}
\affiliation{
$^1$School of Electrical and Electronic Engineering, Nanyang Technological University, 50 Nanyang Avenue, Singapore 639798 \\
$^2$Institute for Digital Molecular Analytics and Science, 59 Nanyang Drive, Singapore 636921
}

\begin{abstract}
Resonances of open non-Hermitian systems are associated with the poles of the system scattering matrix. Perturbations of the system cause these poles to shift in the complex frequency plane. In this work, we introduce a novel method for calculating shifts in scattering matrix poles using generalized Wigner-Smith operators. We link our method to traditional cavity perturbation theory and validate its effectiveness through application to complex photonic networks. Our findings underscore the versatility of generalized Wigner-Smith operators for analyzing a broad spectrum of resonant systems and provides new insight into resonant properties of non-Hermitian systems. 
\end{abstract}

\maketitle

\emph{Introduction}---Resonance is a fundamental physical phenomenon that plays a key role in fields as diverse as quantum mechanics \cite{PhysRevA.95.022117}, structural engineering \cite{Kappos2014}, electromagnetics \cite{gay2002electromagnetic}, and fluid dynamics \cite{zhao2022nonlinear}. In closed systems, resonances are associated with normal modes, which correspond to the orthogonal eigenvectors of the system's Hermitian Hamiltonian. The corresponding real eigenvalues define the spectrum of permissible oscillation frequencies or energy levels \cite{Krasnok2019}.  Eigenvalue problems, however, generally do not admit  analytical solutions. Approximate methods, in which a system is modelled by perturbing a solvable reference problem, are therefore frequently employed. Rayleigh-Schr\"{o}dinger perturbation theory, in which the perturbed solution is written as a series expansion over all unperturbed eigenstates, is perhaps the most famous approach \cite{Schrodinger1926}, however a multitude of alternate theories have been developed, such as Brillouin-Wigner, many-body or lattice perturbation theory \cite{Brillouin1932,PhysRev.136.B896, Capitani2003}.

Although closed systems are useful theoretical ideals, in reality a system interacts with its environment. The coupling of normal modes to external degrees of freedom introduces potential loss channels, which can alter the nature and energy of resonant modes. So-called open, or scattering, systems can no longer be described by a Hermitian Hamiltonian and a coupling operator must be introduced to form an effective non-Hermitian Hamiltonian \cite{Ashida2020}. Alternatively, one can use the closely related scattering matrix, $\mathbf{S}$, which derives from the resolvent of the effective Hamiltonian \cite{RevModPhys.89.015005}. The resonant states of the system, now termed \emph{quasi}-normal modes (QNMs), are then identified from the poles of $\mathbf{S}$ when analytically continued into the complex frequency plane \cite{Kristensen:20,Krasnok2019}. Notably, loss pushes the resonant frequencies off the real axis, resulting in complex-valued eigenfrequencies $\omega_p$, where $\real(\omega_p)$ and $-\imag(\omega_p)/2$ are the resonant frequency and linewidth. 

Open systems are known to possess a number of interesting resonant and scattering phenomena, such as coherent perfect absorption \cite{Baranov2017}, exceptional points \cite{miri2019exceptional}, and bound states in the continuum \cite{Hsu2016}. Moreover, resonances in open systems are susceptible to environmental variations, such as changes in temperature, external fields or material composition, a concept underpinning modern sensors \cite{chen2020optical, RevModPhys.89.035002, Foreman2015b}.  Perturbed non-Hermitian systems can exhibit additional unique behavior, such as anomalous resonance shifts \cite{Ruesink2015, Azeem2021} and nonlinear sensitivity \cite{Parto2020}. Evaluation of resonant modes and their properties therefore remains an important task in, for example, evaluating state lifetimes, laser dynamics and sensor sensitivity. Traditional perturbation theories, however, are not directly applicable to open systems due to mathematical complexities introduced by QNMs \cite{Sternheim1972,Ching1998}. For high quality factor resonances, such as those supported by optical microcavities \cite{Vahala2003}, coupling to the environment is weak allowing resonance shifts to be expressed as changes in stored energy \cite{Waldron1960,bethe1943perturbation}. The validity of such expressions is however limited and they fail to correctly predict changes in linewidths \cite{https://doi.org/10.1002/lpor.201700113}, which is now understood to be a signature of the non-orthogonality of QNMs \cite{PhysRevLett.108.184101, PhysRevLett.113.224101}. In the presence of significant losses, more careful consideration of the normalization of QNMs \cite{https://doi.org/10.1002/lpor.201700113} or alternative mode expansions \cite{PhysRevApplied.11.044018} become necessary.

In this paper, we present a novel formulation for determining resonance shifts due to arbitrary system perturbations based on generalized Wigner-Smith (WS) operators. In the case of high quality factor resonances, we show that these operators are related to a system's stored and dissipated energy, in full agreement with results from traditional perturbation theory. Our theory is tested against numerical simulations of resonant scattering in random photonic networks. In an  accompanying paper \cite{WS_second_paper}, we apply our theory to low quality factor resonances and present further numerical examples.

\emph{Perturbation theory}---An important tool commonly employed in the analysis of scattering systems is the WS time delay matrix
\begin{align}\label{eq:ws-definition}
	\mathbf{Q}_\omega = -i\mathbf{S}^{-1}\partialdiff{\mathbf{S}}{\omega},
\end{align}
where $\partialdiff{}{\omega}$ denotes the partial derivative  with respect to frequency $\omega$ \cite{TEXIER201616, 9142355}. $\mathbf{Q}_\omega$ has several interesting properties relevant to resonant scattering. For unitary systems at real $\omega$, $\mathbf{Q}_\omega$ is Hermitian and its eigenvalues are associated with well-defined time delays experienced by narrow-band, transient system excitations. Evaluated at scattering matrix poles, these time delays coincide with the lifetimes of the associated resonances, demonstrating a link between $\mathbf{Q}_\omega$ and a system's QNMs \cite{DECARVALHO200283}. In electromagnetic theory, it has also been shown that diagonal elements of $\mathbf{Q}_\omega$ can be expressed as energy-like integrals over the extent of the system \cite{10091775}, highlighting the connection to mode volume. 

By itself $\mathbf{Q}_\omega$ lacks the specificity required to efficiently capture localized or parametric system perturbations. For this reason, we also consider the  generalized WS operator $\mathbf{Q}_\alpha$, defined by Eq. (\ref{eq:ws-definition}) with $\omega \rightarrow \alpha$, where $\alpha$ is an arbitrary variable. Operators of this kind have recently attracted  attention in the optical domain as a tool for engineering light in complex scattering environments \cite{PhysRevLett.119.033903, Horodynski2019, Bouchet2021}. Importantly, by virtue of its generalized nature, $\mathbf{Q}_\alpha$  contains sufficient information to predict resonance shifts induced by arbitrary system perturbations. To show this, we consider a generic scattering system described by a scattering matrix $\mathbf{S}$, whose poles are assumed to be of order one.  Let $\alpha$ ($\alpha'$) denote some system parameter, such as the temperature, shape or size of some component of the system, before (after) it is perturbed.  To determine the change in the resonant frequency $\Delta\omega_p = \omega_p' - \omega_p$ that is induced by the perturbation $\Delta\alpha = \alpha' - \alpha$, consider the function $f(\omega, \alpha) = \det[\mathbf{S}^{-1}(\omega, \alpha)]$, whose zeros coincide with the poles of $\mathbf{S}$. Assuming the perturbation is small, a first order, multivariate Taylor expansion of $f$ gives
\begin{align}\label{eq:taylor-series-brief}
	f' = f + (\omega'-\omega) \partialdiff{f}{\omega} + (\alpha' - \alpha)\partialdiff{f}{\alpha},
\end{align}
where $f = f(\omega, \alpha)$ and $f' = f(\omega', \alpha')$. Ideally, we would evaluate Eq. (\ref{eq:taylor-series-brief}) at $\omega = \omega_p$, $\omega' = \omega'_p$ and solve for $\Delta\omega_p$. Some care is required, however, in recasting the result back in terms of $\mathbf{S}$, whose elements diverge at $\omega_p$. This problem can be mitigated by calculating the residue of each term on the right hand side of Eq. (\ref{eq:taylor-series-brief}) at $\omega=\omega_p$, which ultimately yields (see Supplemental Material for more details \cite{supplemental})
\begin{align}\label{eq:pole-shift-main}
	\Delta\omega_p &= i\Delta\alpha\res{\omega_p}{\tr(\mathbf{Q}_\alpha)} = -\Delta\alpha\frac{\res{\omega_p}{\tr(\mathbf{Q}_\alpha)}}{\res{\omega_p}{\tr(\mathbf{Q}_\omega)}},
\end{align}
where $\tr$ denotes the trace operator and $\res{\omega_p}{}$ denotes the residue at the pole $\omega_p$. As can be seen from Eq. (\ref{eq:pole-shift-main}), for a given perturbation $\Delta \alpha$, the pole shift $\Delta \omega_p$ is completely determined by $\res{\omega_p}{\tr(\mathbf{Q}_\alpha)}$, which should be evaluated for the \emph{unperturbed} network. We also observe that the  form of Eq. (\ref{eq:pole-shift-main}) implies the ratio of the residues of the generalized WS operators behaves as a condition number for $\omega_p$ when treated as a function of $\alpha$
\cite{trefethen1997numerical}. Eq. (\ref{eq:pole-shift-main}) is our key result, which, importantly, was derived on purely mathematical grounds, and is therefore applicable to a broad range of physical scenarios.

To gain physical insight into Eq. (\ref{eq:pole-shift-main}), we consider now a class of optical systems composed of arbitrary, finite interior regions coupled to the environment by a finite number of dielectric waveguides. Examples of such systems include fiber Bragg resonators, complex networks and ring resonators \cite{othonos1997fiber,gaio2019nanophotonic, rabus2020ring}. Note that Eq. (\ref{eq:pole-shift-main}) can be written in the  form 
\begin{align}\label{eq:pole-shift-limit}
	\Delta\omega_p = \lim_{\omega \to \omega_p}\left[-\Delta\alpha\frac{\tr(\mathbf{Q}_\alpha)}{\tr(\mathbf{Q}_\omega)}\right],
\end{align}
and so the pole shift is given approximately by the expression within the limit evaluated at $\omega \approx \omega_p$. We assume for simplicity that the system permittivity $\epsilon$ is real and isotropic and the permeability $\mu_0$ is that of free space. Restricting to resonances have high quality factors, such that $\imag(\omega) \ll \real(\omega)$, it is possible to show that \cite{supplemental}
\begin{align}\label{eq:ws-energy-main}
	\mathbf{Q}_\xi& =\left[\mathbf{I} + 2\imag(\omega)\int_\Omega(\epsilon\mathbf{U}^{e} + \mu_0\mathbf{U}^{m})\,\mathrm{d}V\right]^{-1}
	\nonumber\\
	&\quad\quad\bigg[-\int_\Omega\big[\partialdiff{(\omega\epsilon)}{\xi}\mathbf{U}^{e} +\partialdiff{\omega}{\xi}\mu_0\mathbf{U}^{m} \nonumber\\
	&\quad\quad\quad\quad\quad+ 2i\imag(\omega)(\epsilon\mathbf{V}_\xi^{e} +\mu_0\mathbf{V}_\xi^{m})\big]\,\dee V\bigg],
\end{align}
where $\xi$ represents either $\omega$ or $\alpha$ and $\Omega$ denotes the volume occupied by the system with a boundary $\partial\Omega$ perforated only by the coupling waveguides. In Eq. (\ref{eq:ws-energy-main}), $\mathbf{I}$ is the identity matrix and $\mathbf{U}^{e}, \mathbf{U}^{m}, \mathbf{V}^{e}_\xi$, and $\mathbf{V}^{m}_\xi$ are matrices whose $(q,p)$--th elements are given by $U^{e}_{qp} = \frac{1}{4}\mathbf{E}_p\cdot\mathbf{E}^*_q$, $U^{m}_{qp} = \frac{1}{4}\mathbf{H}_p\cdot\mathbf{H}^*_q$, $V^{e}_{\xi,qp} = \frac{1}{4}\partialdiff{\mathbf{E}_p}{\xi}\cdot\mathbf{E}^*_q$, and $V^{m}_{\xi,qp} = \frac{1}{4}\partialdiff{\mathbf{H}_p}{\xi}\cdot\mathbf{H}^*_q$ respectively, where the fields $\mathbf{E}_p$ and $\mathbf{H}_p$ ($\mathbf{E}_q$ and $\mathbf{H}_q$) are those that exist throughout $\Omega$ when the system is illuminated by the $p$--th ($q$--th) incident field. Here, $p$ and $q$  enumerate all modes in all waveguides that couple the system to the environment. 

As discussed in the Supplemental Material \cite{supplemental}, the form of $\mathbf{Q}_\xi$ in Eq. (\ref{eq:ws-energy-main}) is linked to the factorization $\mathbf{Q}_\xi = \mathbf{A}^{-1}\mathbf{B}_\xi$, where $\mathbf{A} = \mathbf{S}^\dagger\mathbf{S}$ and $\mathbf{B}_\xi = -i\mathbf{S}^\dagger\partialdiff{\mathbf{S}}{\xi}$. We shall analyze these factors separately. First, to further simplify matters, we consider the case where the system has a single coupling waveguide supporting a single mode, so that the scattering matrix reduces to an effective scalar reflection coefficient $r_\mathrm{eff}$. Correspondingly, the WS operator $\mathbf{Q}_\xi$ reduces to the scalar quantity $Q_\xi = A^{-1}B_\xi$, where $A = |r_\mathrm{eff}|^2$ and $B_\xi = -ir^*_\mathrm{eff}\partialdiff{r_\mathrm{eff}}{\xi}$. To evaluate these expressions at  complex $\omega$, we recast the problem into one at a real frequency, but with modified material parameters. Introducing $\tilde\epsilon = \epsilon[1 + i\imag(\omega)/\real(\omega)]$ and $\tilde\mu = \mu_0[1 + i\imag(\omega)/\real(\omega)]$, we can show that \cite{supplemental}
\begin{align}\label{eq:A}
	A - 1 = \frac{\real(\omega)}{2} \int_\Omega[\imag(\tilde\epsilon)|\mathbf{E}|^2 + \imag(\tilde\mu)|\mathbf{H}|^2]\,\dee V,
\end{align}
where the fields should now be understood to oscillate with real frequency $\real(\omega)$. Note, the non-zero imaginary parts of $\tilde\epsilon$ and $\tilde\mu$ introduce virtual gain or loss to the system, which is an artifact of the fact that $\omega$ is actually complex. Eq. (\ref{eq:A}) is an energy balance relation, since the integral on the right hand side describes the energy dissipated (gained) within the system by the virtual loss (gain) \cite{Geyi:19}. $B_\omega$ meanwhile is complex  with \cite{supplemental}
\begin{align}\label{eq:omega-energy}
	\real(B_\omega) = -\frac{1}{4}\int_\Omega &\big[\partialdiff{(\real(\omega) \epsilon)}{\real(\omega)}|\mathbf{E}|^2 + \mu_0|\mathbf{H}|^2 \nonumber\\
	&-2\real(\omega)(\imag(\tilde{\epsilon}) \imag(\partialdiff{\mathbf{E}}{\real(\omega)}\cdot \mathbf{E}^*) \nonumber\\
	& + \imag(\tilde{\mu}) \imag(\partialdiff{\mathbf{H}}{\real(\omega)}\cdot \mathbf{H}^*))\big] \,\dee V,
\end{align}
which, up to constant factors, equals the system's stored electromagnetic energy \cite{Geyi:19}. Frequency derivatives in Eq. (\ref{eq:omega-energy}) are necessary to  account for material dispersion. On the other hand, $\imag(B_\omega) = -\partialdiff{A}{\real(\omega)}/2$  describes energy dissipation within the system. Similar expressions have been used to account for material losses in resonant systems \cite{PhysRevA.85.031805, PhysRevA.90.043847}.

Consider now $B_\alpha$ and suppose that the system perturbation modifies the permittivity in a localized region $\Omega_\alpha \subset \Omega$. This might be caused by, for example, a local pressure change or a particle binding to the system. If the  internal field distributions are only weakly affected by the perturbation, then $V^{e}_\alpha = V^{m}_\alpha \approx 0$ and 
\begin{align}\label{eq:ws-numerator}
	B_\alpha \approx  \frac{\omega}{4}\int_\Omega\partialdiff{\epsilon}{\alpha}|\mathbf{E}|^2\dee V.
\end{align}
Recalling the form of the right hand side of Eq. (\ref{eq:pole-shift-limit}), note that the numerator will contain the factor $\Delta\alpha B_\alpha$, which, in light of Eq. (\ref{eq:ws-numerator}), will involve the product $\Delta\alpha \partialdiff{\epsilon}{\alpha}$.  By assumption, outside of $\Omega_\alpha$, $\partialdiff{\epsilon}{\alpha} = 0$, while within $\Omega_\alpha$ we have $\Delta\alpha\partialdiff{\epsilon}{\alpha} = \Delta\epsilon$, where $\Delta\epsilon$ is the change in $\epsilon$. Eq. (\ref{eq:pole-shift-limit}) therefore reduces to 
\begin{align}\label{eq:energy-final}
	\frac{\Delta\omega_p}{\omega_p} = - \frac{1}{B_\omega}\int_{\Omega_\alpha}\Delta \epsilon|\mathbf{E}|^2\dee V.
\end{align}
Since we have assumed high quality factor resonances, the  virtual gain or dissipation will be weak, such that $B_\omega \approx \real(B_\omega)$ and Eq. (\ref{eq:energy-final}) is thus in full agreement with standard cavity perturbation theory \cite{https://doi.org/10.1002/lpor.201700113}. It is important to stress that Eq. (\ref{eq:energy-final}) is only strictly valid for infinite quality factor resonances, but is a reasonable approximation when  loss is weak. In contrast, Eq. (\ref{eq:pole-shift-main}) holds more generally, since no restrictions were made in its derivation. Further analysis of low quality factor resonances can be found in Ref.~\cite{WS_second_paper}.

\emph{Numerical examples}---We now turn to demonstrating the validity of our results with numerical simulations. As an example, we consider photonic networks consisting of randomly connected, single-mode dielectric waveguides, which have recently been investigated as a platform for random lasing \cite{gaio2019nanophotonic,Saxena} and in integrated photonic circuits \cite{Wang2023}. Such systems are relatively simple to model and have non-trivial spectra. An example network, shown in Figure~\ref{fig:network}, was generated by Delaunay triangulation of a random collection of coplanar `internal' nodes (purple). An additional layer of `external' nodes (green) connect to selected internal nodes at the network periphery and serve as entry points, allowing one to define the network scattering matrix $\mathbf{S}$. Given knowledge of the scattering properties of the network's nodes and links, $\mathbf{S}$ and its derivatives can be calculated and thus $\mathbf{Q}_\xi$ evaluated directly \cite{Giacomelli2019}. In the Supplemental Material \cite{supplemental}, we  demonstrate how $\mathbf{Q}_\xi$ can also be found numerically from Eq.~(\ref{eq:ws-energy-main}). 
\begin{figure}[t]
	\centering
	\includegraphics[width=\columnwidth]{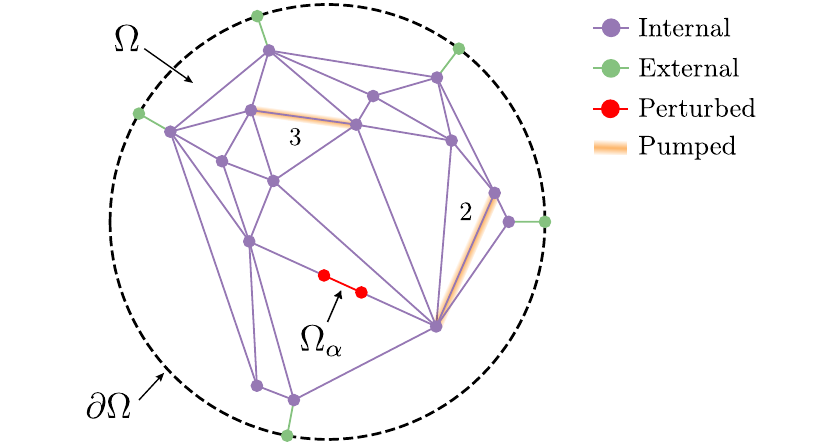}
	\caption{An example random network. Purple nodes and links are `internal', while green `external' elements allow light to enter and exit the network. Red link ($\Omega_{\alpha}$) and highlighted links (2 and 3) correspond to those perturbed in numerical experiments described in the main text.}
	\label{fig:network}
\end{figure}

With our network, we searched for  poles of $\mathbf{S}$ with  frequencies $\real(\omega) \sim 3425\,\mathrm{THz}$, corresponding to a vacuum wavelength of $\sim 550\,\mathrm{nm}$. The links were assumed to be made of glass with refractive index $n$ calculated using a standard Sellmeier equation \cite{Polyanskiy2024}. Dispersion was weak in our example and $n \approx 1.5185$ was approximately constant over the frequency range considered. Propagation of light through each link was described using exponential factors of the form $e^{\pm i\omega nL/c}$, where $L$ is the link length. The average link length in the network was $60\,\mu$m. For simplicity, each internal node was given a randomly generated, frequency-independent scattering matrix drawn from the circular orthogonal ensemble to enforce energy conservation and reciprocity \cite{Byrnes2022}. To perturb the network, we isolated a segment of a randomly chosen link (red `perturbed' segment, $\Omega_\alpha$, in Figure~\ref{fig:network}) and varied its refractive index $n_s$. Specifically, $n_s$ was given by $n_s = n + \Delta n$, where $\Delta n$ was incrementally increased from $10^{-5}$ to $10^{-2}$. Note that the final value $\Delta n = 10^{-2}$ corresponded to a total phase shift in the segment $\phi = \omega \Delta n L /c$ equal to several multiples of $2\pi$. We further introduced two virtual nodes at the ends of the perturbed segment, which were given standard Fresnel scattering matrices based on the refractive index mismatch.
\begin{figure}[t]
	\centering
	\includegraphics[width=\columnwidth]{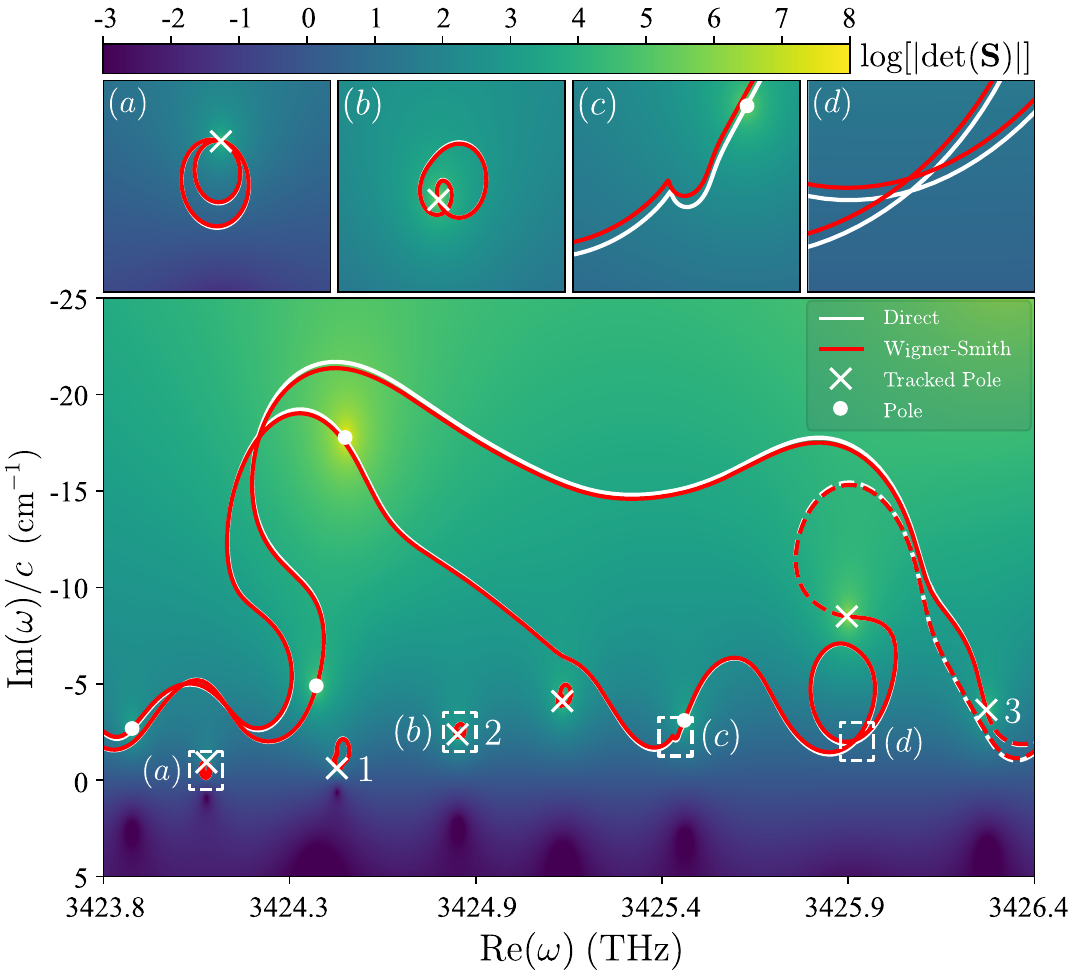}
	\caption{(Main panel) Complex $\omega$ plane showing the trajectories of the network scattering matrix poles upon perturbation of $\Omega_{\alpha}$. Curves (solid for $\Delta n>0$ and dashed $\Delta n<0$) were traced (white) by numerically solving $\det(\mathbf{S}^{-1}) = 0$ and (red) from Eq. (\ref{eq:pole-shift-main}). White crosses (dots) denote the (un)tracked poles. Numbered poles are those described in the numerical pumping experiment. White dashed boxes depict regions shown in (a)--(d).}
	\label{fig:pole-shifts-network}
\end{figure}

Figure \ref{fig:pole-shifts-network} depicts the variation of $\log[|\det(\mathbf{S})|]$ in the complex $\omega$ plane for the unperturbed network, i.e., $\Delta n = 0$. Bright regions, marked with white dots and crosses, correspond to the poles of the unperturbed network scattering matrix. Figures~\ref{fig:pole-shifts-network}(a)-(d) show detailed plots of the smaller regions bounded by the white dashed boxes in the main panel. The positions of a subset of the poles, specifically those marked with white crosses, were tracked as $\Delta n$ was increased. The trajectories followed are shown by the solid red and white lines emanating from the initial pole positions (crosses). White lines were traced by numerically solving $\det(\mathbf{S}^{-1}) = 0$ for each value of $\Delta n$. The red lines, on the other hand, were traced using Eq.~(\ref{eq:pole-shift-main}) with $\alpha = \Delta n$ to determine the pole shifts at each step (see Supplemental Material \cite{supplemental} for details of the calculation of $\mathbf{Q}_\alpha$). As can be seen, the WS theory agrees excellently with the direct numerical solutions. Some discrepancies, however, can be seen, e.g., near the top of the main panel. In this region, it was found that  computed values for $\res{\omega_p}{\tr(\mathbf{Q}_\alpha)}$ were relatively large, implying that the pole shifts were more sensitive to the perturbation. Accordingly, a smaller perturbation step size is requried to restore validity of Eq.~(\ref{eq:taylor-series-brief}). Dashed red and white lines emanating from the crosses on the right hand side of the figure show the  pole trajectories for $\Delta n <0$ and show further agreement between the two methods.

Figure~\ref{fig:pole-shifts-network} shows several interesting features. First, note that the half space $\imag(\omega) > 0$ contains dark regions for which $\det(\mathbf{S}) = 0$. These zeros can be pushed to the real axis by introducing loss to the system, whereby they correspond to coherent perfect absorption modes \cite{PhysRevLett.105.053901}. Note also that the poles in our data exhibit two distinct types of trajectories: some follow long, meandering paths, while others revolve around localized, closed loops. Figures~\ref{fig:pole-shifts-network}(a)-(b) in particular demonstrate the latter type. It is important to realize that $\mathbf{S}$ is quasiperiodic in $\Delta n$ since for specific values of $\Delta n$ the additional propagation phase acquired in $\Omega_\alpha$ will be an exact multiple of $2\pi$. At these values, the only difference between the perturbed and unperturbed network will be the weak Fresnel matrices at the ends of the perturbed link segment, the effect of which is negligible for small $\Delta n$. Poles following loop trajectories must therefore return to their original positions, while poles following open trajectories must pass through the positions of other poles of the unperturbed network (white dots in Figure~\ref{fig:pole-shifts-network}). 

\begin{figure}[t]
	\centering
	\includegraphics[width=\columnwidth]{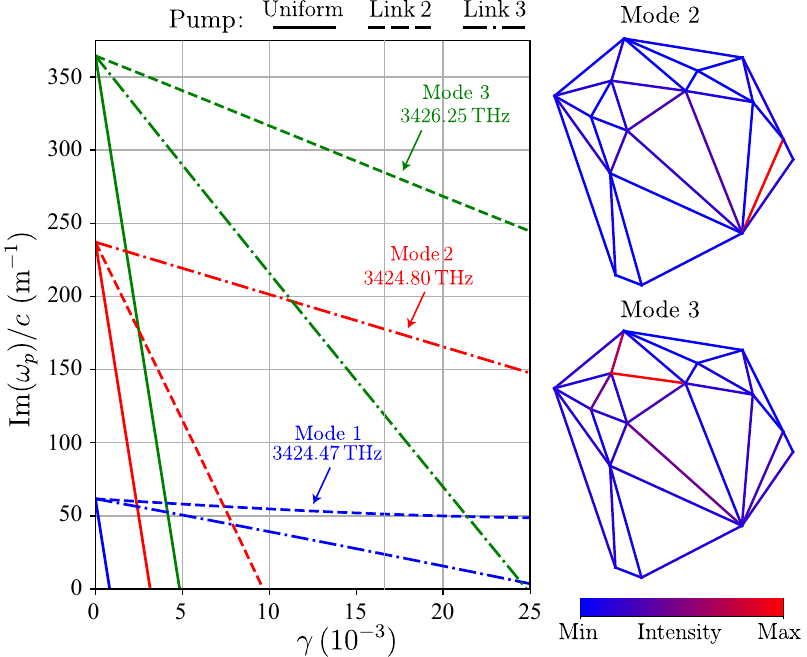}
	\caption{(left) Imaginary parts of selected poles as different network links are pumped (see Figure~\ref{fig:network}). Colors and line styles distinguish different modes and pumping methods respectively. Spatial intensity profiles of modes 2 (top right) and 3 (bottom right).}
	\label{fig:pump}
\end{figure}
As a further application of our theory, we next considered adding gain to (or `pumping') our network which can lead to random lasing \cite{Saxena}. Mathematically, gain can be introduced to the $j$'th link by setting its refractive index to $n_j = \real(n_j) - i\gamma_j$, where $\real(n_j)$ is calculated as before and $\gamma_j > 0$ is a gain coefficient taken, for simplicity, to be frequency independent over the pumping bandwidth. Adding gain causes poles to drift towards the real axis and the pole that reaches the axis first typically determines the dominant lasing frequency \cite{Andreasen2010}. Single mode lasing at arbitrary resonant frequencies is achievable by shaping the pump profile in accordance with the spatial profile of the resonant mode \cite{PhysRevLett.109.033903, Bachelard2014}. Similar selective lasing can be predicted using our theory by considering the collection of generalized WS operators associated with the gain coefficients. Specifically, for any pole $\omega_p$ and for $j$, we calculate $\res{\omega_p}{\tr(\mathbf{Q}_{\gamma_j})}$, where $\mathbf{Q}_{\gamma_j}$ is the operator associated with adding gain $\gamma_j$ to \emph{only} the $j$'th link. In light of Eq.~(\ref{eq:pole-shift-main}), these values predict how the pole will shift when different links are pumped, thus revealing which should be pumped to optimally shift $\omega_p$ towards threshold.

Figure~\ref{fig:pump} shows the results of the described numerical pumping experiment. The imaginary parts of three arbitrarily selected poles (distinguished by color and corresponding to the numbered poles in Figure \ref{fig:pole-shifts-network}) were tracked in response to three alternate pumping methods (distinguished by line style). Solid lines track the poles when all links were pumped uniformly, in which case the narrowest resonance, mode 1 (at 3424.47 THz), reaches the lasing threshold first. Alternatively, the dashed and dot-dashed lines track the poles when the links highlighted (and denoted $2$ and $3$ respectively) in Figure~\ref{fig:network}, were pumped in isolation. In particular, links 2 and 3 correspond to those predicted to optimally shift modes 2 (at 3424.8 THz) and 3 (at 3426.25 THz) respectively towards the real axis. This predicted behaviour is evident in Figure~\ref{fig:pump}, with pumping of link 2 bringing mode 2 to lasing threshold first (and similarly for mode/link 3). We also calculated the spatial intensity profiles of modes 2 and 3 across the network, which are shown on the right hand side of Figure~\ref{fig:pump}. Notably, the intensity of mode 2 (3) is strongly peaked across link 2 (3), confirming the wisdom that the optimal pump profile should conform to the mode's spatial distribution. Interestingly, however, the use of the generalized WS operator eliminated the need to explicitly calculate the mode distribution in determining the pump profile. Finally, we note that a similar strategy of selectively introducing loss to specific links could produce coherent perfect absorption of desired modes.  

\emph{Conclusion}---In this work we have presented a novel method for calculating resonance shifts in perturbed open systems using generalized WS operators. Our work reveals the connection between generalized WS operators and resonant properties in non-Hermitian systems and further highlights their utility. Our perturbation theory is based on general complex analytic arguments and is therefore applicable to a broad range of scenarios. Furthermore, we demonstrated our results reduce to  traditional perturbation formulas for high quality resonances. We have verified our theory numerically by tracking pole shifts caused by refractive index perturbations in a complex photonic network, and in a spatially selective pumping experiment. Our work provides a novel way to analyze scattering resonances, which may be of use in cavity or nanostructure design \cite{Granchi2023} and future sensing technologies. 

\begin{acknowledgments}
	N.B. was supported by Singapore Ministry of Education Academic Research Fund (Tier 1) Grant RG66/23. M.R.F. was supported by funding from the Institute for Digital
	Molecular Analytics and Science (IDMxS) under the Singapore Ministry of Education Research Centres of Excellence scheme and by Nanyang Technological University Grant SUG:022824-00001.
\end{acknowledgments}


\providecommand{\noopsort}[1]{}\providecommand{\singleletter}[1]{#1}%

\clearpage

\setcounter{equation}{0}
\setcounter{figure}{0}
\renewcommand\theequation{S\arabic{equation}}
\renewcommand\thefigure{S\arabic{figure}}

\let\oldthebibliography=\thebibliography
\let\oldendthebibliography=\endthebibliography
\newenvironment{thesuppbibliography}[1]{
	\oldthebibliography{#1}
	\setcounter{enumiv}{1}                       
}{\oldendthebibliography}

\title{Supplemental material for ``Perturbing scattering resonances in non-Hermitian systems: a generalized Wigner-Smith operator formulation''} 

\maketitle

\onecolumngrid
\section{\label{sec:perturbation}Derivation of the pole shift equation} 
As discussed in the main text, it is helpful to consider the function 
\begin{align}
	f(\omega, \alpha) = \det[\mathbf{S}^{-1}(\omega, \alpha)],
\end{align}
where $\mathbf{S}$ is the scattering matrix, $\omega$ is the complex frequency, and $\alpha$ is a system parameter. Importantly, if $\omega_p$ is a pole of $\mathbf{S}$, then $f(\omega_p, \alpha) = 0$. We assume that all poles and zeros are of order one. Suppose that the system is perturbed so that $\alpha$ changes value to $\alpha' = \alpha + \Delta\alpha$ and let $C$ be an arbitrary contour in the complex frequency plane that encircles $\omega_p$ and no other poles or zeros of $\mathbf{S}$. If $\omega \in C$, expanding $f$ about $(\omega, \alpha)$ we have
\begin{align}\label{eq:taylor-full}
	f(\omega_p', \alpha') = f(\omega, \alpha) + (\omega'_p-\omega) &\partialdiff{f}{\omega}(\omega, \alpha) + \Delta\alpha\partialdiff{f}{\alpha}(\omega, \alpha) + \cdots,
\end{align}
where $\omega'_p$ is the shifted pole for the perturbed system. The higher order terms in Eq. (\ref{eq:taylor-full}) will contain factors of the form $(\omega_p'-\omega)^{n_1}\Delta\alpha^{n_2}$, where $n_1 + n_2 \geq 2$. We shall neglect these terms on the grounds that the perturbation $\Delta\alpha$ and the corresponding pole shift $\Delta\omega_p$ are assumed to be small. By taking $C$ to be sufficiently close to $\omega_p$, we can assume that $\omega \approx \omega_p$ and thus $\omega'_p - \omega \approx \omega'_p - \omega_p = \Delta\omega_p$. Therefore, $(\omega_p'-\omega)^{n_1}\Delta\alpha^{n_2} \approx \Delta\omega_p^{n_1}\Delta\alpha^{n_2}$ and since all terms for which $n_1 + n_2 \geq 2$ are the products of multiple small quantities, we neglect them.

Since $\omega'_p$ is a resonant frequency of the perturbed system, $f(\omega_p', \alpha')=0$. Also, since $f$ is non-zero on $C$, we can divide each term in Eq. (\ref{eq:taylor-full}) by $f(\omega, \alpha)$ to give (dropping functional dependencies on $\omega$ and $\alpha$ in our notation for clarity)
\begin{align}\label{eq:taylor-short}
	0 = 1 + (\omega'_p - \omega)\frac{\partialdiff{f}{\omega}}{f} + \Delta\alpha\frac{\partialdiff{f}{\alpha}}{f}.
\end{align}
Letting $\xi$ stand in place of $\omega$ and $\alpha$ and using standard results from matrix calculus, the logarithmic derivative of $f$ is given by \cite{SHjørungnes_2011}\begin{align}\label{eq:f-combo-derivative}
	\frac{\partialdiff{f}{\xi}}{f} = \tr(\mathbf{S}\partialdiff{\mathbf{S}^{-1}}{\xi}) = -\tr(\mathbf{S}^{-1}\partialdiff{\mathbf{S}}{\xi}).
\end{align}
The final equality in Eq. (\ref{eq:f-combo-derivative}) follows from the fact that 
\begin{align}\label{eq:matrix-diff-inverse}
	\partialdiff{\mathbf{S}^{-1}}{\xi} = -\mathbf{S}^{-1}(\partialdiff{\mathbf{S}}{\xi})\mathbf{S}^{-1}
\end{align}
and the cyclic invariance of the trace operator. 

Ideally, we would evaluate Eq. (\ref{eq:taylor-short}) at $\omega = \omega_p$ and solve the resulting equation for $\Delta\omega_p$. Since $f$ has a zero at $\omega_p$, however, the function $\partialdiff{f}{\xi}/f$ has a pole at $\omega_p$. To avoid problems associated with divergences, we  proceed by calculating the residue of each term in Eq. (\ref{eq:taylor-short}) at $\omega_p$. This can be done by dividing each term by $2\pi i$ and integrating over the contour $C$. Clearly the unity term has zero residue. For the term containing $\partialdiff{f}{\alpha}/f$, it follows from Eq. (\ref{eq:f-combo-derivative}) and the definition of the Wigner-Smith operator that
\begin{align}\label{eq:residue-alpha}
	\frac{\Delta\alpha}{2\pi i}\oint_C \frac{\partialdiff{f}{\alpha}}{f}\dee\omega &= -i\Delta\alpha\frac{1}{2\pi i}\oint_C \tr(-i\mathbf{S}^{-1}\partialdiff{\mathbf{S}}{\alpha})\dee\omega
	= -i\Delta\alpha\res{\omega_p}{\tr(\mathbf{Q}_\alpha)}.
\end{align}
The residue of the term containing $\partialdiff{f}{\omega}/f$ can be evaluated using the argument principle \cite{Sahlfors1979complex}. Since, by construction, $f$ only has a single zero (at $\omega_p$) and no poles within $C$, it follows that
\begin{align}\label{eq:argument}
	\res{\omega_p}\frac{\partialdiff{f}{\omega}}{f} = 1.
\end{align}
Next, we make use of the fact that if the functions $g$ and $h$ are such that $g$ is holomorphic at $\omega_p$ and $h$ has a simple pole at $\omega_p$, then
\begin{align}
	\res{\omega_p}{gh} = g(\omega_p)\res{\omega_p}{h}.
\end{align}
Setting $g(\omega) = \omega'_p - \omega$ and $h = \partialdiff{f}{\omega}/f$, we obtain
\begin{align}
	\frac{1}{2\pi i}\oint_C (\omega'_p - \omega)\frac{\partialdiff{f}{\omega}}{f} \mathrm{d}\omega = \omega'_p - \omega_p = \Delta\omega_p.
\end{align}
Having found the residue of each term in Eq. (\ref{eq:taylor-short}), the resulting equation can now be solved for $\Delta\omega_p$ to give 
\begin{align}
	\Delta\omega_p &= i\Delta\alpha\res{\omega_p}{\tr(\mathbf{Q}_\alpha)},
\end{align}
which is the first equality of Eq. (\ref{eq:pole-shift-main}) in the main text.

The second equality of Eq. (\ref{eq:pole-shift-main}) in the main text follows from the fact that
\begin{align}
	\res{\omega_p}{\tr(\mathbf{Q}_{\omega})} = i,
\end{align}
which is a straightforward consequence of Eqs. (\ref{eq:f-combo-derivative}) and (\ref{eq:argument}) since
\begin{align}
	i = i\frac{1}{2\pi i}\oint_C\frac{\partialdiff{f}{\omega}}{f} \dee\omega &=  \frac{1}{2\pi i}\oint_C\tr(-i\mathbf{S}^{-1}\partialdiff{\mathbf{S}}{\omega}) \dee\omega = \res{\omega_p}{\tr(\mathbf{Q}_{\omega})}.
\end{align}
Finally, since $i = -1/i$, we have
\begin{align}
	\Delta\omega_p = i\Delta\alpha\res{\omega_p}{\tr(\mathbf{Q}_\alpha)} = -\Delta\alpha\frac{\res{\omega_p}{\tr(\mathbf{Q}_\alpha)}}{i} = -\Delta\alpha\frac{\res{\omega_p}{\tr(\mathbf{Q}_\alpha)}}{\res{\omega_p}{\tr(\mathbf{Q}_\omega)}},
\end{align}
which completes the derivation.

\section{\label{sec:wigner-smith}Derivation of the volume integral form of the Wigner-Smith operator} 

In this section we present a derivation of Eq. (\ref{eq:ws-energy-main}) in the main text, which expresses the Wigner-Smith operators in terms of volume integrals over the system.

An important observation is that the Wigner-Smith operator can be factorized as $\mathbf{Q}_\xi = \mathbf{A}^{-1}\mathbf{B}_\xi$, where $\mathbf{A} = \mathbf{S}^\dagger\mathbf{S}$ and $\mathbf{B}_{\xi} = -i\mathbf{S}^\dagger\partialdiff{\mathbf{S}}{\xi}$. We proceed by deriving volume integral expressions for $\mathbf{A}$ and $\mathbf{B}_\xi$ separately and then combining the results to obtain an expression for $\mathbf{Q}_\xi$. As shall be shown, the former can be derived from a generalized Poynting theorem, while the latter can be derived from an additional energy balance relation.

Suppose that all fields are time harmonic with an implicit $e^{-i\omega t}$ dependence. The permittivity $\epsilon$ is assumed to be a real valued scalar and the permeability $\mu_0$ is that of free-space. We begin by determining a expression for $\mathbf{A}$. Let $\Omega$ be a large surface that encapsulates the system and let $\partial\Omega$ denote its boundary. Suppose that light is able to enter and exit the system via a finite number of waveguides that perforate $\partial\Omega$. Let $\mathbf{E}_p$ and $\mathbf{H}_p$ ($\mathbf{E}_q$ and $\mathbf{H}_q$) denote the fields throughout the system that arise due to illuminating the system by the $p$--th ($q$--th) incident field, where $p$ ($q$) enumerates all of the modes in all of the waveguides. An integral version of Poynting's theorem for time harmonic fields gives \cite{SGeyi:19}
\begin{align}
	\frac{1}{2}\int_{\partial \Omega}(\mathbf{E}_p \times \mathbf{H}^*_q)\cdot\hat{\mathbf{n}}\,\mathrm{d}A = \frac{i}{2}\int_\Omega(\omega\mu_0\mathbf{H}_p \cdot \mathbf{H}^*_q - \omega^*\epsilon\mathbf{E}_p  \cdot \mathbf{E}^*_q)\,\mathrm{d}V ,\label{eq:poynting}
\end{align}
where $\hat{\mathbf{n}}$ is an outward-pointing unit normal vector to the surface $\partial\Omega$. 

Consider first the integral on the left hand side of Eq. (\ref{eq:poynting}), which can be written as a sum of integrals over the waveguide cross sections. We shall assume for simplicity that these waveguides only support a single mode, but extension to multiple modes is straightforward. Around the cross section $\partial\Omega_m$, where $\partial\Omega$ meets the the $m$--th waveguide, we define a local Cartesian coordinate system where $\hat{\mathbf{z}}$ points out of the system along the waveguide axis and $\hat{\mathbf{x}}$ and $\hat{\mathbf{y}}$ lie within $\partial\Omega_m$, which is assumed to be at $z=0$. Let $\mathbf{E}_{mp}$ and $\mathbf{H}_{mp}$ ($\mathbf{E}_{mq}$ and $\mathbf{H}_{mq}$) denote the fields within the $m$'th waveguide that arise due to illuminating the system via the $p$--th ($q$--th) waveguide. We assume that the fields within the waveguide have the form
\begin{align}
	\mathbf{E}_{mp} &= \delta_{mp}\mathbf{e}_m^-e^{-i\beta_m z} + S_{mp}\mathbf{e}_m^+e^{i\beta_m z},\label{eq:E-guide}\\
	\mathbf{H}_{mp} &= \delta_{mp}\mathbf{h}_m^-e^{-i\beta_m z} + S_{mp}\mathbf{h}_m^+e^{i\beta_m z},\label{eq:H-guide}
\end{align}
where $\beta_m$ is the waveguide propagation constant, $\delta_{mp}$ is a Kronecker delta, $S_{mp}$ is the $(m,p)$--th element of the scattering matrix, and $\mathbf{e}_m^+$, $\mathbf{e}_m^-$, $\mathbf{h}_m^+$, and $\mathbf{h}_m^-$ are the vector profiles of the electric and magnetic fields, which are functions only of the transverse spatial coordinates $x$ and $y$. The propagation constant can be expressed as $\beta_m = n^{\mathrm{eff}}_{m}k_0$, where $n^\mathrm{eff}_{m} = \sqrt{\epsilon^\mathrm{eff}_{m}/\epsilon_0}$ is an effective refractive index, and $k_0 = \omega/c$ is the vacuum wavenumber, where $c$ is the speed of light in vacuum. Using Eqs. (\ref{eq:E-guide}) and (\ref{eq:H-guide}), we have		
\begin{align}
	\begin{split}
		\frac{1}{2}&\int_{\partial\Omega_m}(\mathbf{E}_{mp} \times \mathbf{H}^*_{mq})\cdot\hat{\mathbf{n}}\,\mathrm{d}A\\
		&= \frac{1}{2}\int_{\partial\Omega_m}[(\delta_{mp}\mathbf{e}_{m,t}^- + S_{mp}\mathbf{e}_{m,t}^+) \times (\delta_{mq}\mathbf{h}_{m,t}^{-*} + S^*_{mq}\mathbf{h}_{m,t}^{+*})]\cdot\hat{\mathbf{z}}\,\mathrm{d}A\\
		&=\frac{1}{2}(\delta_{mp}\delta_{mq} - \delta_{mp}S^*_{mq} + \delta_{mq}S_{mp} - S_{mp}S^*_{mq})\int_{\partial\Omega_m}(\mathbf{e}_{m,t}\times\mathbf{h}_{m,t}^*)\cdot\hat{\mathbf{z}}\,\mathrm{d}A,
		\label{eq:poynting-cross}
	\end{split}
\end{align}
where $\mathbf{e}^\pm_{m,t}$ and $\mathbf{h}^\pm_{m,t}$ are the transverse parts of $\mathbf{e}^\pm_{m}$ and $\mathbf{h}^\pm_{m}$. Note that we have adopted the convention in which $\mathbf{e}_{m,t} = \mathbf{e}_{m,t}^- = \mathbf{e}_{m,t}^+$ and $\mathbf{h}_{m,t} = \mathbf{h}_{m,t}^- = -\mathbf{h}_{m,t}^+$, which can be done without loss of generality \cite{Ssnyder2012optical}. 

The integral on the right side of the final equality of Eq. (\ref{eq:poynting-cross}) warrants some discussion. Ideally, we would assert that the integral is equal to unity by some appropriate mode normalization. For materials with loss or gain, however, this cannot be assumed to be the case. In our case, even though $\epsilon$ is real, we are still faced with virtual loss or gain in virtue of of the fact that $\omega$ is complex (see Section \ref{sec:complex}). Since, however, we are only concerned with high quality factor resonances for which $\imag(\omega)/\real(\omega) \ll 1$, this effect is weak and we are permitted to treat $\mathbf{e}_{m,t}$ and $\mathbf{h}_{m,t}$ as though the waveguides were lossless, allowing us to assert \cite{Ssnyder2012optical}
\begin{align}\label{eq:normalize}
	\frac{1}{2}\int_{\partial\Omega_m}(\mathbf{e}_{m,t}\times\mathbf{h}_{m,t}^*)\cdot\hat{\mathbf{z}}\,\mathrm{d}A = 1.
\end{align}
If the external waveguides supported multiple modes, we would also be permitted to assume mode orthogonality in the sense that the integral in Eq. (\ref{eq:normalize}) would vanish if $\mathbf{e}_t$ and $\mathbf{h}_t$ were associated with two different modes. 

Armed with Eq. (\ref{eq:normalize}), summing Eq. (\ref{eq:poynting-cross}) over $m$ gives the integral over all of $\partial\Omega$, which is given by 
\begin{align}
	\frac{1}{2}\int_{\partial\Omega}(\mathbf{E}_{p} \times \mathbf{H}^*_{q})\cdot \hat{\mathbf{n}}\,\mathrm{d}A = (\mathbf{I} - \mathbf{S}^\dagger\mathbf{S} + \mathbf{S} - \mathbf{S}^\dagger)_{qp},
\end{align}
where $\mathbf{I}$ is the identity matrix and the subscript $qp$ indicates that we are looking at the $(q,p)$--th element of the matrix expression within the parentheses. In order to ease notation, let
\begin{align}
	U^{e}_{qp} &= \frac{1}{4}\mathbf{E}_p  \cdot \mathbf{E}^*_q,\\
	U^{m}_{qp} &= \frac{1}{4}\mathbf{H}_p  \cdot \mathbf{H}^*_q,
\end{align}
and let $\mathbf{U}^{e}$ and $\mathbf{U}^{m}$ denote the matrices whose $(q,p)$--th elements are $U^{e}_{qp}$ and $U^{m}_{qp}$ respectively. Eq. (\ref{eq:poynting}) can then be written as the matrix equation
\begin{align}
	\mathbf{I} - \mathbf{S}^\dagger\mathbf{S} + \mathbf{S} - \mathbf{S}^\dagger = 2i\int_\Omega(\omega\mu_0\mathbf{U}^{m} - \omega^*\epsilon\mathbf{U}^{e})\,\mathrm{d}V.\label{eq:poynting-final}
\end{align}
In order to extract $\mathbf{S}^\dagger\mathbf{S}$ from Eq. (\ref{eq:poynting-final}), we take the Hermtiain part of both sides. The Hermitian and skew Hermitian parts of a matrix $\mathbf{M}$, which we denote by $\her(\mathbf{M})$ and $\sher(\mathbf{M})$ respectively, are defined by 
\begin{align}
	\mathrm{Her}(\mathbf{M}) &= \frac{\mathbf{M} + \mathbf{M}^\dagger}{2},\\
	\mathrm{SHer}(\mathbf{M}) &= \frac{\mathbf{M} - \mathbf{M}^\dagger}{2}.
\end{align}
Taking the Hermitian part eliminates the skew Hermitian matrix $\mathbf{S} - \mathbf{S}^{\dagger}$ from the left side of Eq. (\ref{eq:poynting-final}), leaving only the Hermitian matrix $\mathbf{I} - \mathbf{S}^\dagger\mathbf{S}$. Finally, we find $\mathbf{A}$ to be \begin{align}\label{eq:sdaggers}		\mathbf{A} = \mathbf{S}^\dagger\mathbf{S} = \mathbf{I} + 2\imag(\omega)\int_\Omega(\epsilon\mathbf{U}^{e} + \mu_0\mathbf{U}^{m})\,\mathrm{d}V.
\end{align}
Note that if $\imag(\omega) = 0$, the system is truly lossless and the integral in  Eq. (\ref{eq:sdaggers}) vanishes, recovering the usual unitarity relation for $\mathbf{S}$.

We now turn to the problem of determining an integral equation for $\mathbf{B}_\xi$. For our starting point, it can be shown in general that, on the basis of energy conservation (see, for example, Ref. \cite{SGeyi:19}),
\begin{align}\label{eq:energycons}
	\begin{split}
		\frac{i}{4}\int_{\partial\Omega} (\mathbf{E}_p \times &\partialdiff{\mathbf{H}^*_q}{\xi} \pm \partialdiff{\mathbf{E}_p}{\xi} \times \mathbf{H}^*_q)\cdot \hat{\mathbf{n}}\,\mathrm{d}A\\ 
		= \int_\Omega [&\partialdiff{( \omega^*\epsilon)}{\xi}U^{e}_{qp} \mp \partialdiff{\omega}{\xi}\mu_0U^{m}_{qp}+\omega^*\epsilon(V^{e*}_{\xi,pq} \pm V^{e}_{\xi,qp})-\omega\mu_0(V^{m*}_{\xi,pq} \pm V^{m}_{\xi,qp})]\,\mathrm{d}V,
	\end{split}
\end{align}
where we define
\begin{align}
	V^{e}_{\xi,qp} &= \frac{1}{4}\partialdiff{\mathbf{E}_p}{\xi} \cdot \mathbf{E}^*_q,\\
	V^{m}_{\xi,qp} &= \frac{1}{4}\partialdiff{\mathbf{H}_p}{\xi} \cdot \mathbf{H}^*_q.
\end{align}
Note that Eq. (\ref{eq:energycons}) is in fact a pair of equations as one may choose either the upper or lower set of plus and minus signs. The surface integral on the left hand side of Eq. (\ref{eq:energycons}) can be evaluated using similar steps to those in deriving Eq. (\ref{eq:poynting-cross}), taking appropriate derivatives and conjugates of Eqs. (\ref{eq:E-guide}) and (\ref{eq:H-guide}) as required. For the $m$--th waveguide, we find
\begin{align}
	\begin{split}\label{eq:big-ws-intermediate}
		&\frac{1}{4}\int_{\partial\Omega_m} (\mathbf{E}_{mp} \times \partialdiff{\mathbf{H}^*_{mq}}{\xi} \pm \partialdiff{\mathbf{E}_{mp}}{\xi} \times \mathbf{H}^*_{mq})\cdot \hat{\mathbf{z}}\,\mathrm{d}A\\
		&= \frac{1}{2}(-\delta_{mp}\partialdiff{S^*_{mq}}{\xi} - S_{mp}\partialdiff{S^*_{mq}}{\xi}\pm \delta_{mq}\partialdiff{S_{mp}}{\xi}  \mp S^*_{mq}\partialdiff{S_{mp}}{\xi})\\
		&\quad+\frac{1}{2}(\delta_{mp}\delta_{mq} - \delta_{mp}S^*_{mq} + \delta_{mq}S_{mp} - S_{mp}S^*_{mq})\int_{\partial\Omega_m}(\mathbf{e}_{m,t} \times \partialdiff{\mathbf{h}^*_{m,t}}{\xi} \pm \partialdiff{\mathbf{e}_{m,t}}{\xi} \times \mathbf{h}^*_{m,t})\cdot\hat{\mathbf{z}}\,\mathrm{d}A.
	\end{split}
\end{align}
Assuming that the transverse mode profiles $\mathbf{e}_{m,t}$ and $\mathbf{h}_{m,t}$ only weakly depend on $\xi$, the integral in the final term of Eq. (\ref{eq:big-ws-intermediate}) vanishes. Summing the resulting equation over $m$ gives
\begin{align}
	&\frac{i}{4}\int_{\partial\Omega} (\mathbf{E}_{p} \times \partialdiff{\mathbf{H}^*_{q}}{\xi} \pm (\partialdiff{\mathbf{E}_{p}}{\xi}) \times \mathbf{H}^*_{q})\cdot \hat{\mathbf{n}}\,\mathrm{d}A = \frac{1}{2}[-i\partialdiff{\mathbf{S}^\dagger}{\xi} - i\partialdiff{\mathbf{S}^\dagger}{\xi}\mathbf{S} \pm i\partialdiff{\mathbf{S}}{\xi} \mp i\mathbf{S}^\dagger\partialdiff{\mathbf{S}}{\xi}]_{qp}
\end{align}
and, as before, Eq. (\ref{eq:energycons}) can therefore be written as the matrix equation
\begin{align}\label{eq:wsonway}
	\begin{split}
		\frac{1}{2}&(-i\partialdiff{\mathbf{S}^\dagger}{\xi} - i\partialdiff{\mathbf{S}^\dagger}{\xi}\mathbf{S} \pm i\partialdiff{\mathbf{S}}{\xi} \mp i\mathbf{S}^\dagger\partialdiff{\mathbf{S}}{\xi})\\
		&= \int_\Omega [\partialdiff{( \omega^*\epsilon)}{\xi}\mathbf{U}^{e} \mp \partialdiff{\omega}{\xi}\mu_0\mathbf{U}^{m} + \omega^*\epsilon(\mathbf{V}_\xi^{e \dagger} \pm \mathbf{V}_\xi^{e})-\omega\mu_0(\mathbf{V}_\xi^{ m \dagger} \pm \mathbf{V}_\xi^{m})]\,\mathrm{d}V,
	\end{split}
\end{align}
where $\mathbf{V}_\xi^{e}$ ($\mathbf{V}_\xi^{m}$) is the matrix whose $(q,p)$--th element is $V^{e}_{\xi,qp}$ ($V^{m}_{\xi,qp}$). To isolate $-i\mathbf{S}^\dagger\partialdiff{\mathbf{S}}{\xi}$, we take the skew Hermitian parts of both sides of Eq. (\ref{eq:wsonway}) with the upper set of plus and minus signs and the Hermitian parts of both sides of Eq. (\ref{eq:wsonway}) with the lower set of plus and minus signs. This gives the pair of equations
\begin{align}\label{eq:her-part}
	\begin{split}
		\sher(&-i\mathbf{S}^\dagger\partialdiff{\mathbf{S}}{\xi})\\
		&= -i\int_\Omega[\partialdiff{(\imag(\omega)\epsilon)}{\xi}\mathbf{U}^{e} +\partialdiff{\imag(\omega)}{\xi}\mu_0\mathbf{U}^{m} -2i\imag(\omega)(\epsilon\her(\mathbf{V}_\xi^{e}) +\mu_0\her(\mathbf{V}_\xi^{m}))]\,\dee V,
	\end{split}
\end{align}
\begin{align}\label{eq:sher-part}
	\begin{split}
		\her(&-i\mathbf{S}^\dagger\partialdiff{\mathbf{S}}{\xi})\\
		&= -\int_\Omega[\partialdiff{(\real(\omega)\epsilon)}{\xi}\mathbf{U}^{e} +\partialdiff{\real(\omega)}{\xi}\mu_0\mathbf{U}^{m}
		-2i\imag(\omega)(\epsilon\sher(\mathbf{V}_\xi^{e}) +\mu_0\sher(\mathbf{V}_\xi^{m}))]\,\dee V.
	\end{split}
\end{align}
Next, since, by definition,
\begin{align}
	\mathbf{B}_\xi = -i\mathbf{S}^\dagger\partialdiff{\mathbf{S}}{\xi} = \her(-i\mathbf{S}\partialdiff{\mathbf{S}}{\xi}) +\sher(-i\mathbf{S}^\dagger\partialdiff{\mathbf{S}}{\xi}),
\end{align}
Eqs. (\ref{eq:her-part}) and (\ref{eq:sher-part}) can be combined to give
\begin{align}\label{eq:QH-fin}
	\begin{split}
		\mathbf{B}_\xi = &-\int_\Omega[\partialdiff{(\omega\epsilon)}{\xi}\mathbf{U}^{e} +\partialdiff{\omega}{\xi}\mu_0\mathbf{U}^{m} + 2i\imag(\omega)(\epsilon\mathbf{V}_\xi^{e} +\mu_0\mathbf{V}_\xi^{m})]\,\dee V
		.
	\end{split}
\end{align}
Finally, combining $\mathbf{A}$ and $\mathbf{B}_\xi$, we arrive at
\begin{align}\label{eq:ws-final}
	\begin{split}
		\mathbf{Q}_\xi = &\bigg(\mathbf{I} + 2\imag(\omega)\int_\Omega(\epsilon\mathbf{U}^{e} + \mu_0\mathbf{U}^{m})\,\mathrm{d}V\bigg)^{-1}\\
		&\bigg(-\int_\Omega[\partialdiff{(\omega\epsilon)}{\xi}\mathbf{U}^{e} +\partialdiff{\omega}{\xi}\mu_0\mathbf{U}^{m} + 2i\imag(\omega)(\epsilon\mathbf{V}_\xi^{e} +\mu_0\mathbf{V}_\xi^{m})]\,\dee V\bigg),
	\end{split}
\end{align}
which is Eq. (\ref{eq:ws-energy-main}) in the main text.

\section{\label{sec:complex}Physical interpretation of complex frequency and derivation of energy interpretation of Wigner-Smith operators}
In this section we discuss the physical interpretation of complex frequencies and derive Eqs. (\ref{eq:A}) and (\ref{eq:omega-energy}) in the main text.

Since $\omega = \real{(\omega)}[1 +i\imag{(\omega)}/\real{(\omega)}]$, Maxwell's curl equations throughout the system can be written as
\begin{align}
	\nabla \times \mathbf{E} & =i\omega\mu_0\mathbf{H} =i\real{(\omega)}\tilde{\mu}\mathbf{H}, \label{eq:max-E}\\
	\nabla \times \mathbf{H} &= -i\omega\epsilon\mathbf{E}  = -i\real{(\omega)} \tilde\epsilon\mathbf{E}\label{eq:max-H},
\end{align}
where $\tilde{\epsilon}=\epsilon[1 +i\imag{(\omega)}/\real{(\omega)}]$ and $\tilde{\mu}=\mu_0[1 +i\imag{(\omega)}/\real{(\omega)}]$. This demonstrates a physical equivalence between two different points of view. On the one hand, we may think of the system as having \emph{real} material parameters $\epsilon$ and $\mu_0$, but supporting waves at a \emph{complex} frequency $\omega$. On the other hand, we may instead imagine that the waves oscillate at the \emph{real} frequency $\real(\omega)$, but in a system with \emph{complex} material parameters $\tilde\epsilon$ and $\tilde\mu$. Since Maxwell's equations are the same in either case, the two viewpoints are physically equivalent. The latter, however, is perhaps more familiar, and the imaginary parts of the permittivity and permeability are well understood as representing loss or gain. A wave possessing a complex frequency therefore also exhibits loss or gain, depending on the sign of $\imag(\omega)$.

Using the definitions of $\tilde\epsilon$ and $\tilde\mu$ in Eq. (\ref{eq:sdaggers}), it is straightforward to show that
\begin{align}
	\mathbf{A} = \mathbf{I} + 2\real(\omega)\int_\Omega[\imag(\tilde\epsilon)\mathbf{U}^{e} + \imag(\tilde\mu)\mathbf{U}^{m}]\,\mathrm{d}V,
\end{align}
which, when restricted to the single mode example discussed in the main text, which enforces the transformations $\mathbf{I} \to 1$, $\mathbf{U}^e \to \frac{1}{4}|\mathbf{E}|^2$ and $\mathbf{U}^m \to \frac{1}{4}|\mathbf{H}|^2$, reduces to Eq. (\ref{eq:A}) in the main text. Similalrly, in terms of $\tilde\epsilon$ and $\tilde\mu$, Eq. (\ref{eq:QH-fin}) becomes 
\begin{align}\label{eq:ws-final-modified}
	\begin{split}
		\mathbf{B}_\xi = -\int_\Omega[\partialdiff{(\real(\omega)\tilde\epsilon)}{\xi}\mathbf{U}^{e} +\partialdiff{(\real(\omega)\tilde\mu)}{\xi}\mathbf{U}^{m} + 2i\real(\omega)(\imag(\tilde\epsilon)\mathbf{V}_\xi^{e} +\imag(\tilde\mu)\mathbf{V}_\xi^{m})]\,\dee V.
	\end{split}
\end{align}
Restricting again to a single mode, where now $\mathbf{V}^e_\xi \to \frac{1}{4}\partialdiff{\mathbf{E}}{\xi}\cdot\mathbf{E}^*$ and $\mathbf{V}^m_\xi = \frac{1}{4}\partialdiff{\mathbf{H}}{\xi}\cdot\mathbf{H}^*$, and taking the real part of Eq. (\ref{eq:ws-final-modified}), we find
\begin{align}\label{eq:real-part}
	\begin{split}
		\real({B}_\xi) = 
		-\frac{1}{4}\int_\Omega[&\partialdiff{(\real(\omega)\epsilon)}{\xi}|\mathbf{E}|^2 +\partialdiff{\real(\omega)\mu_0}{\xi}|\mathbf{H}|^2\\
		&- 2\real(\omega)(\imag(\tilde\epsilon)\imag(\mathbf\partialdiff{\xi}{\mathbf{E}}\cdot\mathbf{E}^*) +\imag(\tilde\mu)\imag(\partialdiff{\mathbf{H}}{\xi}\cdot\mathbf{H}^*))]\,\dee V.
	\end{split}
\end{align}
Next, note that for any holomorphic function $f$ of a complex variable $z$ we have $\partialdiff{f}{z} = \partialdiff{f}{\real(z)}$ \cite{Sahlfors1979complex}. Assuming that all functions in Eq. (\ref{eq:real-part}) that need to be differentiated are complex differentiable with respect to $\omega$, we can set $\xi = \omega$ in Eq. (\ref{eq:real-part}) and make the substitution $\partialdiff{}{\omega} \to \partialdiff{}{\real(\omega)}$ to obtain
\begin{align}\label{eq:real-part2}
	\begin{split}
		\real({B}_\omega) = 
		-\frac{1}{4}\int_\Omega[&\partialdiff{(\real(\omega)\epsilon)}{\real(\omega)}|\mathbf{E}|^2 +\mu_0|\mathbf{H}|^2\\
		&- 2\real(\omega)(\imag(\tilde\epsilon)\imag(\mathbf\partialdiff{\real(\omega)}{\mathbf{E}}\cdot\mathbf{E}^*) +\imag(\tilde\mu)\imag(\partialdiff{\mathbf{H}}{\real(\omega)}\cdot\mathbf{H}^*))]\,\dee V,
	\end{split}
\end{align}
which is Eq. (\ref{eq:omega-energy}) in the main text. Finally, taking the imaginary of part Eq. (\ref{eq:ws-final-modified}) for $\xi = \omega$ gives
\begin{align}\label{eq:imag-part}
	\begin{split}
		\imag({B}_\omega) = 
		-\frac{1}{4}\int_\Omega[&\partialdiff{(\real(\omega)\imag(\tilde\epsilon))}{\real(\omega)}|\mathbf{E}|^2 +\partialdiff{(\real(\omega)\imag(\tilde\mu))}{\real(\omega)}|\mathbf{H}|^2\\
		&+ 2\real(\omega)(\imag(\tilde\epsilon)\real(\mathbf\partialdiff{\real(\omega)}{\mathbf{E}}\cdot\mathbf{E}^*) +\imag(\tilde\mu)\real(\partialdiff{\mathbf{H}}{\real(\omega)}\cdot\mathbf{H}^*))]\,\dee V.
	\end{split}
\end{align}
Since 
\begin{align}
	\real(\partialdiff{\mathbf{E}}{\omega}\cdot\mathbf{E}^*) = \frac{1}{2}(\partialdiff{\mathbf{E}}{\omega}\cdot\mathbf{E}^* + \partialdiff{\mathbf{E}^*}{\omega}\cdot\mathbf{E}) = \frac{1}{2}\partialdiff{|\mathbf{E}|^2}{\omega}
\end{align}
and, similarly, $\real(\partialdiff{\mathbf{H}}{\omega}\cdot\mathbf{H}^*) = \frac{1}{2}\partialdiff{|\mathbf{H}|^2}{\omega}$, Eq. (\ref{eq:imag-part}) can be simplified to give
\begin{align}
	\imag({B}_\omega) = 
	-\frac{1}{2}\partialdiff{}{\real(\omega)}\Bigg(\frac{\real(\omega)}{2}\int_\Omega[\imag(\tilde\epsilon)|\mathbf{E}|^2 +\imag(\tilde\mu)|\mathbf{H}|^2]\,\dee V\Bigg) = -\frac{1}{2}\partialdiff{A}{\real(\omega)},
\end{align}
which shows that $\imag(B_\omega)$ is related to energy dissipation within the system.

\section{\label{sec:energy} Evaluation of the Wigner-Smith volume integrals for complex photonic networks }

\begin{figure}[H]
	\centering
	\includegraphics[scale=1.1]{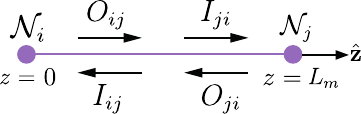}
	\caption{Geometry of an internal network link.}
	\label{fig:link}
\end{figure}
In this section we show how to evaluate the volume integrals appearing in Eq. (\ref{eq:ws-final}) for a complex photonic network. The structure and properties of these networks are discussed in the main text. 

The first integral in Eq. (\ref{eq:ws-final}) is
\begin{align}\label{eq:first-integral}
	\mathbf{A} - \mathbf{I} = 2\imag(\omega)\int_\Omega(\epsilon\mathbf{U}^{e} + \mu_0\mathbf{U}^{m})\,\mathrm{d}V.
\end{align}
Integrating over $\Omega$ requires integrating over all of the  $N_\mathrm{node}$ nodes and $N_\mathrm{link}$ links in the network as well as the fields in the surrounding space. We can decompose $\Omega$ into a union of subsets 
\begin{align}
	\Omega = \Omega^\mathrm{space}\cup\Omega^\mathrm{link}\cup\Omega^\mathrm{node} = \Omega^\mathrm{space}\cup(\Omega_1^\mathrm{link}\cup\hdots\cup \Omega_{N_\mathrm{link}}^\mathrm{link})\cup(\Omega_1^\mathrm{node}\cup\hdots\cup \Omega_{N_\mathrm{node}}^\mathrm{node}),
\end{align}
where $\Omega_m^{\mathrm{link}}$ is the volume occupied by the $m$--th link, $\Omega_m^{\mathrm{node}}$ is the volume occupied by the $m$--th node, and $\Omega^\mathrm{space}$ is the volume occupied by the remaining space around the links and nodes. Assuming the system does not support leaky modes, the fields in the space around the links and nodes will be evanescent, decaying away from the network components. If necessary, we may expand the volumes $\Omega_m^\mathrm{link}$ and $\Omega_m^\mathrm{node}$ beyond the physical extent of the network components so as to contain these evanescent fields up to a point where they have sufficiently decayed and no longer significantly contribute to the integrals. The integral over the remaining space $\Omega^\mathrm{space}$ can then be be neglected. 

Consider evaluating the integral in Eq. (\ref{eq:first-integral}) over $\Omega^{\mathrm{link}}_m$ and suppose that this link connects the $i$--th and $j$--th nodes $\mathcal{N}_i$ and $\mathcal{N}_j$. A diagram of the link is given in Figure~\ref{fig:link}. We define a local Cartesian coordinate system for the link, similar to that used in Section \ref{sec:wigner-smith}. The unit vector $\hat{\mathbf{z}}$ is assumed to be parallel to the link's axis and points into the link from $\mathcal{N}_i$ at $z=0$. The link extends to $z= L_m$, where it meets $\mathcal{N}_j$ such that $L_m$ is the length of the link. To keep our calculations relatively simple, we assume that each link is weakly guiding with an effective refractive index $n_m^\mathrm{eff} = c\sqrt{\epsilon^\mathrm{eff}_m\mu_0}$, uniform in the $z$ direction and approximately constant over transverse cross sections. We also assume that each link supports a single mode, which, given the weak guiding assumption, has negligible $z$ component and transverse fields that satisfy \cite{Ssnyder2012optical}
\begin{align}
	\mathbf{h}_{m,t} = Y^\mathrm{eff}_m\hat{\mathbf{z}}\times\mathbf{e}_{m,t},
\end{align}
where $Y^\mathrm{eff}_m = \sqrt{\epsilon^\mathrm{eff}_m/\mu_0}$ is the link's admittance. We can now derive individual normalization equations for $\mathbf{e}_{m,t}$ and $\mathbf{h}_{m,t}$. Let $\partial \Omega^\mathrm{link}_m$ be a transverse cross section of the link (for any value of $z$). Recalling Eq. (\ref{eq:normalize}), we have
\begin{align}
	\begin{split}
		1 = \frac{1}{2}\int_{\partial \Omega^\mathrm{link}_m}(\mathbf{e}_{m,t}\times\mathbf{h}^*_{m,t})\cdot\hat{\mathbf{z}}\,\dee A &= \frac{Y^\mathrm{eff}_m}{2}\int_{\partial \Omega^\mathrm{link}_m}[\mathbf{e}_{m,t}\times(\hat{\mathbf{z}}\times\mathbf{e}_{m,t}^*)]\cdot\hat{\mathbf{z}}\,\dee A\\
		&= \frac{Y^\mathrm{eff}_m}{2}\int_{\partial \Omega^\mathrm{link}_m}\mathbf{e}_{m,t}\cdot\mathbf{e}^*_{m,t}\,\dee A,
	\end{split}
\end{align}
from which it follows that
\begin{align}\label{eq:e-norm}
	\frac{1}{2}\int_{\partial \Omega^\mathrm{link}_m}\mathbf{e}_{m,t}\cdot\mathbf{e}^*_{m,t}\,\dee A = \frac{1}{Y^\mathrm{eff}_m}.
\end{align}
Similarly,
\begin{align}\label{eq:h-norm}
	\begin{split}
		\frac{1}{2}\int_{\partial \Omega^\mathrm{link}_m}\mathbf{h}_{m,t}\cdot\mathbf{h}^*_{m,t}\,\dee A &= \frac{(Y^\mathrm{eff}_m)^2}{2}\int_{\partial \Omega^\mathrm{link}_m}(\hat{\mathbf{z}}\times\mathbf{e}_{m,t})\cdot(\hat{\mathbf{z}}\times\mathbf{e}^*_{m,t})\,\dee A\\
		&= \frac{(Y_m^{^\mathrm{eff}})^2}{2}\int_{\partial \Omega^\mathrm{link}_m}\mathbf{e}_{m,t}\cdot\mathbf{e}^*_{m,t}\,\dee A= Y^\mathrm{eff}_m.
	\end{split}
\end{align}

The fields within the link that arise due to illuminating the system through the $p$--th external link are given by
\begin{align}
	\mathbf{E}_{mp} = (O_{ij,p}e^{i\beta_m z} + I_{ij,p}e^{-i\beta_m z})\mathbf{e}_{m,t},\label{eq:E-internal}\\
	\mathbf{H}_{mp} = (O_{ij,p}e^{i\beta_m z} - I_{ij,p}e^{-i\beta_m z})\mathbf{h}_{m,t}\label{eq:H-internal},
\end{align}
where $O_{ij,p}$ denotes the field amplitude that enters the link at $z=0$ from $\mathcal{N}_i$ and propagates towards $\mathcal{N}_j$, and $I_{ij,p}$ is the field amplitude that exits the link at $z=0$, entering $\mathcal{N}_i$, having arrived from $\mathcal{N}_j$. Note that the link index $m$ defines the pair of indices $ij$ uniquely. Integrating the field products over $\Omega^\mathrm{link}$ amounts to employing the normalization conditions Eqs. (\ref{eq:e-norm}) and (\ref{eq:h-norm}) to handle the transverse coordinates and integrating the exponential functions with respect to $z$. Summing the results over $m$, we ultimately arrive at
\begin{align}\label{eq:integral-one}
	\begin{split}
		2\imag(\omega)\int_{\Omega^\mathrm{link}}(\epsilon U^{e}_{qp} + \mu_0 U^{m}_{qp})\,\dee V= \sum_{m} [&O_{ij,p}O^*_{ij,q}(e^{-2\imag(\omega)n^\mathrm{eff}_mL_m/c_0} -1)\\
		&+ I_{ij,p}I^*_{ij,q}(1 - e^{2\imag(\omega)n^\mathrm{eff}_mL_m/c_0})],
	\end{split}
\end{align}
which expresses the integral in terms of the internal field components.

\begin{figure}[t]
	\centering
	\includegraphics[scale=1.1]{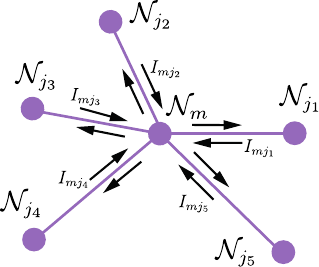}
	\caption{Geometry of an internal network node.}
	\label{fig:node}
\end{figure}
Consider next evaluating the integral in Eq. (\ref{eq:first-integral}) over $\Omega^\mathrm{node}_m$. Suppose that the $m$--th node $\mathcal{N}_m$ is connected to a collection of other nodes $\mathcal{N}_{j_1}, \mathcal{N}_{j_2}, \hdots$ with indices $j_1, j_2, \hdots$ by a collection of links. The geometry of the node is depicted in Figure~\ref{fig:node}. In general, evaluating the integral requires a model for the fields within the node. For simplicity, we shall instead evaluate the integral indirectly. Recall that the integral in question originated from Eq. (\ref{eq:sdaggers}), which was derived from a generalized Poynting theorem over the extent of the network. We can instead apply Poynting's theorem to $\Omega^\mathrm{node}_m$ to obtain a similar equation to Eq. (\ref{eq:sdaggers}) for the node integral in terms of the node scattering matrix $\mathbf{S}_m$. Unlike in our derivation of Eq. (\ref{eq:sdaggers}), however, we cannot use Eqs. (\ref{eq:E-guide}) and (\ref{eq:H-guide}) for the fields, but must instead use expressions for the internal network fields, similar to those in Eqs. (\ref{eq:E-internal}) and (\ref{eq:H-internal}). Repeating the derivation of Eq. (\ref{eq:sdaggers}) with the correct field expressions, we obtain
\begin{align}\label{eq:integral-two}
	2\imag(\omega)\int_{\Omega^\mathrm{node}_m}(\epsilon U^{e}_{qp} + \mu_0 U^{m}_{qp})\,\mathrm{d}V = \mathbf{i}_{mq}^\dagger (\mathbf{S}_m^\dagger\mathbf{S}_m - \mathbf{I}) \mathbf{i}_{mp},
\end{align}
where $\mathbf{i}_{mp} = (I_{mj_1,p}, I_{mj_2,p}, \hdots)^\mathrm{T}$ is a vector containing components of all of the fields that are incident upon the node when the network is illuminated via the $p$--th mode. $\mathbf{i}_{mq} = (I_{mj_1,q}, I_{mj_2,q}, \hdots)^\mathrm{T}$ is defined similarly. Summing the result of Eq. (\ref{eq:integral-two}) over $m$ gives the total integral over $\Omega^\mathrm{node}$.

Consider now the second integral that appears in Eq. (\ref{eq:ws-final}), i.e.
\begin{align}\label{eq:integral-2nd}
	\mathbf{B}_\xi = -\int_\Omega[\partialdiff{(\omega\epsilon)}{\xi}\mathbf{U}^{e} +\partialdiff{\omega}{\xi}\mu_0\mathbf{U}^{m} + 2i\imag(\omega)(\epsilon\mathbf{V}_\xi^{e} +\mu_0\mathbf{V}_\xi^{m})]\,\dee V.\end{align}
As before, we can assume that the integral over $\Omega^\mathrm{space}$ is negligible. Evaluating the integral over the network's links can be done using analogous steps to those used in deriving Eq. (\ref{eq:integral-one}). The algebra is somewhat lengthy and there is little additional insight to be gained from a detailed presentation. The final result is
\begin{align}\label{eq:integral-three}
	\begin{split}
		-\int_{\Omega^\mathrm{link}}&[\partialdiff{(\omega\epsilon)}{\xi}U^{e}_{qp} +\partialdiff{\omega}{\xi}\mu_0U^{m}_{qp} + 2i\imag(\omega)(\epsilon V_{\xi,qp}^{e} +\mu_0V_{\xi,qp}^{m})]\,\dee V\\
		=& \sum_m\Big(\partialdiff{(n^\mathrm{eff}_mk_0)}{\xi}L_m(O_{ij,p}O^*_{ij,q}e^{-2\imag(\omega)n^\mathrm{eff}_mL_m/c_0} + I_{ij,p}I^*_{ij,q}e^{2\imag(\omega)n^\mathrm{eff}_mL_m/c_0})\\
		&+ \frac{\imag(\omega)}{2\real(\omega)}\frac{\partialdiff{n^\mathrm{eff}_m}{\xi}}{n^\mathrm{eff}_m}[O_{ij,p}I^*_{ij,q}(e^{2i\real(\omega)n^\mathrm{eff}_mL_m/c_0} -1) + I_{ij,p}O^*_{ij,q}(1 - e^{-2i\real(\omega)n^\mathrm{eff}_mL_m/c_0})]\\
		&+ i[\partialdiff{O_{ij,p}}{\xi}O^*_{ij,q}(1 - e^{-2\imag(\omega)n^\mathrm{eff}_mL_m/c_0}) + \partialdiff{I_{ij,p}}{\xi}I^*_{ij,q}(e^{2\imag(\omega)n^\mathrm{eff}_mL_m/c_0}-1)]\Big).
	\end{split}
\end{align}
Evaluating the integral over the network's nodes can also be done indirectly by repeating the derivation of Eq. (\ref{eq:QH-fin}), but over $\Omega_m^{\mathrm{node}}$ using the internal fields at the node boundary. This time we obtain
\begin{align}\label{eq:integral-four}
	\begin{split}
		-\int_{\Omega^\mathrm{node}_m}[\partialdiff{(\omega\epsilon)}{\xi}U^{e}_{qp} +\partialdiff{\omega}{\xi}\mu_0U^{m}_{qp} &+ 2i\imag(\omega)(\epsilon V_{\xi,qp}^{e} +\mu_0V_{\xi,qp}^{m})]\,\dee V\\
		&= \mathbf{i}^\dagger_{mq}(-i\mathbf{S}^\dagger_m\partialdiff{\mathbf{S}_m}{\xi})\mathbf{i}_{mp} + i\mathbf{i}^\dagger_{mq}(\mathbf{I}-\mathbf{S}^\dagger_m\mathbf{S}_m)\partialdiff{\mathbf{i}_{mp}}{\xi},
	\end{split}
\end{align}
which can be summed over $m$ to give the total integral over $\Omega^\mathrm{node}$. Finally, combining the results from  Eqs. (\ref{eq:integral-one}), (\ref{eq:integral-two}), (\ref{eq:integral-three}) and (\ref{eq:integral-four}) allows us to compute the Wigner-Smith matrix $\mathbf{Q}_\xi$.

\providecommand{\noopsort}[1]{}\providecommand{\singleletter}[1]{#1}%
\begin{thesuppbibliography}{4}%
	\makeatletter
	\providecommand \@ifxundefined [1]{%
		\@ifx{#1\undefined}
	}%
	\providecommand \@ifnum [1]{%
		\ifnum #1\expandafter \@firstoftwo
		\else \expandafter \@secondoftwo
		\fi
	}%
	\providecommand \@ifx [1]{%
		\ifx #1\expandafter \@firstoftwo
		\else \expandafter \@secondoftwo
		\fi
	}%
	\providecommand \natexlab [1]{#1}%
	\providecommand \enquote  [1]{``#1''}%
	\providecommand \bibnamefont  [1]{#1}%
	\providecommand \bibfnamefont [1]{#1}%
	\providecommand \citenamefont [1]{#1}%
	\providecommand \href@noop [0]{\@secondoftwo}%
	\providecommand \href [0]{\begingroup \@sanitize@url \@href}%
	\providecommand \@href[1]{\@@startlink{#1}\@@href}%
	\providecommand \@@href[1]{\endgroup#1\@@endlink}%
	\providecommand \@sanitize@url [0]{\catcode `\\12\catcode `\$12\catcode
		`\&12\catcode `\#12\catcode `\^12\catcode `\_12\catcode `\%12\relax}%
	\providecommand \@@startlink[1]{}%
	\providecommand \@@endlink[0]{}%
	\providecommand \url  [0]{\begingroup\@sanitize@url \@url }%
	\providecommand \@url [1]{\endgroup\@href {#1}{\urlprefix }}%
	\providecommand \urlprefix  [0]{URL }%
	\providecommand \Eprint [0]{\href }%
	\providecommand \doibase [0]{https://doi.org/}%
	\providecommand \selectlanguage [0]{\@gobble}%
	\providecommand \bibinfo  [0]{\@secondoftwo}%
	\providecommand \bibfield  [0]{\@secondoftwo}%
	\providecommand \translation [1]{[#1]}%
	\providecommand \BibitemOpen [0]{}%
	\providecommand \bibitemStop [0]{}%
	\providecommand \bibitemNoStop [0]{.\EOS\space}%
	\providecommand \EOS [0]{\spacefactor3000\relax}%
	\providecommand \BibitemShut  [1]{\csname bibitem#1\endcsname}%
	\let\auto@bib@innerbib\@empty
	\bibitem [{\citenamefont {Hjørungnes}(2011)}]{SHjørungnes_2011}%
	\BibitemOpen
	\bibfield  {author} {\bibinfo {author} {\bibfnamefont {A.}~\bibnamefont
			{Hjørungnes}},\ }\href {https://doi.org/10.1017/CBO9780511921490} {\emph
		{\bibinfo {title} {Complex-Valued Matrix Derivatives: With Applications in
				Signal Processing and Communications}}}\ (\bibinfo  {publisher} {Cambridge
		University Press},\ \bibinfo {year} {2011})\BibitemShut {NoStop}%
	\bibitem [{\citenamefont {Ahlfors}(1979)}]{Sahlfors1979complex}%
	\BibitemOpen
	\bibfield  {author} {\bibinfo {author} {\bibfnamefont {L.}~\bibnamefont
			{Ahlfors}},\ }\href@noop {} {\emph {\bibinfo {title} {Complex Analysis: An
				Introduction to The Theory of Analytic Functions of One Complex Variable}}}\
	(\bibinfo  {publisher} {McGraw-Hill Education},\ \bibinfo {year}
	{1979})\BibitemShut {NoStop}%
	\bibitem [{\citenamefont {Geyi}(2019)}]{SGeyi:19}%
	\BibitemOpen
	\bibfield  {author} {\bibinfo {author} {\bibfnamefont {W.}~\bibnamefont
			{Geyi}},\ }\bibfield  {title} {\bibinfo {title} {Stored electromagnetic field
			energies in general materials},\ }\href
	{https://doi.org/10.1364/JOSAB.36.000917} {\bibfield  {journal} {\bibinfo
			{journal} {J. Opt. Soc. Am. B}\ }\textbf {\bibinfo {volume} {36}},\ \bibinfo
		{pages} {917–925} (\bibinfo {year} {2019})}\BibitemShut {NoStop}%
	\bibitem [{\citenamefont {Snyder}\ and\ \citenamefont
		{Love}(2012)}]{Ssnyder2012optical}%
	\BibitemOpen
	\bibfield  {author} {\bibinfo {author}{\bibfnamefont {A.}~\bibnamefont
			{Snyder}}\ and\ \bibinfo {author} {\bibfnamefont {J.}~\bibnamefont {Love}},\
	}\href {https://doi.org/10.1007/978-1-4613-2813-1} {\emph {\bibinfo {title}
			{Optical Waveguide Theory}}}\ (\bibinfo  {publisher} {Springer New York},\
	\bibinfo {year} {2012})\BibitemShut {NoStop}%
\end{thesuppbibliography}%

\end{document}